\documentclass[fp,twocolumn]{jpsj3}
\usepackage{txfonts}
\usepackage{bm}

\usepackage{color}
\title{Parallelized Stochastic Cutoff Method for Long-Range Interacting Systems}
\author{Eishin Endo$^1$\thanks{E-mail : eishin@solid.apph.tohoku.ac.jp}, Yuta Toga$^1$, and Munetaka Sasaki$^2$}
\inst{$^1$Department of Applied Physics, Tohoku University, Sendai 980-8579, Japan \\
$^2$Faculty of Engineering, Kanagawa University, Yokohama 221-8686, Japan }

\abst{
We present a method of parallelizing the stochastic cutoff (SCO) method, 
which is a Monte-Carlo method for long-range interacting systems. 
After interactions are eliminated by the SCO method, we subdivide a lattice 
into noninteracting interpenetrating sublattices. This subdivision enables us 
to parallelize the Monte-Carlo calculation in the SCO method. Such subdivision is found 
by numerically solving the vertex coloring of a graph created by the SCO method. 
We use an algorithm proposed by Kuhn and Wattenhofer to solve the vertex coloring 
by parallel computation. This method was applied to a two-dimensional magnetic 
dipolar system on an $L\times L$ square lattice to examine its parallelization 
efficiency. The result showed that, in the case of $L=2304$, the speed of computation
increased about $102$ times by parallel computation with $288$ processors. 
}

\begin{document}
\maketitle

\section{Introduction}

It is widely recognized that the recent trend in computational physics 
is parallel computing with a large number of computational resources. This recognition 
is supported by the fact that all of the top 100 supercomputers released in November 2014 
consist of more than ten thousand cores.~\cite{top500} 
Furthermore, computation with a graphics processing unit (GPU) has been a hot topic in recent 
years~\cite{GPU01, GPU02, GPU03, GPU04, GPU05, GPU06, GPU07, GPU08, GPU09} because it enables us 
to perform massively parallel computing at a fraction of the cost. To fully utilize these parallel 
architectures, the development of efficient parallel algorithms is indispensable. Such parallel 
algorithms are required particularly in long-range interacting systems because of their high 
computational cost.

One example in which the parallelization of computation has been successfully achieved 
is the molecular dynamic (MD) method (see Refs.~\citen{Heffelfinger00,Larsson11}, and references therein). 
If there are only short-range forces, parallel computations are performed 
by dividing the simulation box into cubic domains and assigning each domain to 
each processor~\cite{Fincham89}. If the system involves long-range forces 
such as the Coulomb ones, long-range forces for all the molecules are 
efficiently calculated with ${\cal O}(N \log N)$ or ${\cal O}(N)$ computational time 
($N$ is the number of molecules) by sophisticated methods such as 
the Barnes-Hut tree algorithm,~\cite{Appel85,Barnes86,Makino90} 
first multipole method,~\cite{Greengard88,Carrier88} 
and particle mesh Ewald method.~\cite{PME01,PME02} 
Because these methods can be parallelized, it is possible to perform parallel computations 
in the MD methods even in the presence of long-range forces. 

The main reason why parallel computations in the MD methods are relatively simple 
lies in their simultaneous feature. In the MD methods, forces acting 
on all the molecules are calculated at the beginning of each step, and new positions of 
molecules in the next time step are determined simultaneously using forces calculated 
in advance. In contrast, in a normal Monte-Carlo (MC) method, elements such as particles and spins 
are moved one at a time. This sequential feature of the MC method makes parallelization difficult. 
If the system involves only short-range interactions, it is still possible to perform parallel 
computations in the MC method. For example, MC simulations in lattice systems can be 
parallelized by a checkerboard decomposition.~\cite{LandauBinder_checkerboard} 
Even in off-lattice systems, parallel computations are still possible by 
a spatial decomposition method.~\cite{HeffelfingerLewitt96,Uhlherr02,RenOrkoulas07,Sadigh12}
The spatial-decomposition technique is also used in the kinetic (event-driven) MC method in short-range 
interacting systems to parallelize the 
computation.~\cite{Lubachevsky87,Korniss99,Martinez08,Arampatzis12} 
However, such a spatial-decomposition method does not work in long-range interacting systems 
because all the elements interact with each other. Furthermore, the above-mentioned efficient algorithms 
used in the MD method to calculate long-range forces do not work in the MC method 
because these methods calculate long-range forces (and potentials) {\it for all the elements} at once. 
In the MC methods, long-range forces calculated for all the elements become invalid 
after a part of the elements are updated because of the sequential feature of the MC method. 
Therefore, it is difficult for long-range interacting systems to perform parallel 
computations in a normal (and most widely applicable) single-update MC method.

To overcome this difficulty, we utilize the stochastic cutoff (SCO) method.~\cite{SCO,Sasaki10,SCO_book} 
The SCO method is a Monte-Carlo method for long-range interacting lattice systems. 
The basic idea of the method is to switch long-range interactions $V_{ij}$ stochastically 
to either zero or a pseudointeraction ${\bar V}_{ij}$ using the Stochastic Potential 
Switching (SPS) algorithm~\cite{Mak05,MakSharma07}. The SPS algorithm enables us to switch 
the potentials with the detailed balance condition strictly satisfied. Therefore, 
the SCO method does not involve any approximation. Fukui and Todo have 
developed an efficient MC method based on a similar strategy using different 
pseudointeractions and different way of switching interactions~\cite{FukuiTodo09}. 
Because most of the distant and weak interactions are eliminated by being switched 
to zero, the SCO method markedly reduces the number of interactions and computational 
time in long-range interacting systems. For example, in a two-dimensional magnetic 
dipolar system, to which we will apply our MC method later, 
the number of potentials per spin and the computational time for a single-spin update 
are reduced from ${\cal O}(N)$ to ${\cal O}(1)$.

This reduction of potentials also makes it possible to subdivide the lattice into noninteracting 
interpenetrating sublattices, i.e., so that the elements on a single sublattice 
do not interact with each other. This subdivision enables us to parallelize the computation. 
However, one problem is that there is no trivial way of finding such a subdivision. 
In the case of Ising models on a square lattice with nearest-neighbouring interactions, 
the checkerboard decomposition is responsible for it. In contrast, there is no trivial 
subdivision in the this case because interactions are stochastically switched. 
To resolve this problem, we numerically solve the vertex coloring on a graph created by 
the potential switching procedure. This computation is performed in a parallel fashion 
using an algorithm proposed by Kuhn and Wattenhofer.~\cite{KuhnWattenhofer06,book_KW}

The paper is organized as follows: In Sect.~\ref{sec:method}, we briefly explain the SCO method 
and describe the parallel computation of the vertex coloring, which is a key feature of the present method. 
In Sect.~\ref{sec:result}, we show the results obtained by applying the present method to a two-dimensional
magnetic dipolar system. In Sect.~\ref{sec:conclusions}, we give our conclusions.

\section{Methods}
\label{sec:method}
\subsection{Stochastic Cutoff (SCO) Method}
\label{subsec:SCO}
In this subsection, we briefly explain the SCO method. We consider a system with pairwise long-range 
interactions described by a Hamiltonian ${\cal H}=\sum_{i<j}{V}_{ij}(\bm{S}_i,\bm{S}_j)$, 
where $\bm{S}_i$ is a variable associated with the $i$-th element of the system. In the SCO method, 
$V_{ij}$ is stochastically switched to either $0$ or a pseudointeraction $\bar{V}_{ij}$ as
\begin{equation}
V_{ij}(\bm{S}_i,\bm{S}_j)=
\begin{cases}
0& prob.:P_{ij}(\bm{S}_i,\bm{S}_j),\\
\overline{V}_{ij}(\bm{S}_i,\bm{S}_j) & prob.:1-P_{ij}(\bm{S}_i,\bm{S}_j).
\label{sps}
\end{cases}
\end{equation}
The probability $P_{ij}$ and the pseudointeraction $\overline{V}_{ij}$ are given by
\begin{equation}
P_{ij}(\bm{S}_i,\bm{S}_j) = \exp[\beta(V_{ij}(\bm{S}_i,\bm{S}_j)- V_{ij}^{\text{max}})],
\label{eqn:Pij}
\end{equation}
\begin{equation}
\overline{V}_{ij}(\bm{S}_i,\bm{S}_j) = V_{ij}(\bm{S}_i,\bm{S}_j)-\beta^{-1}\log[1-P_{ij}(\bm{S}_i,\bm{S}_j)],
\end{equation}
where $\beta$ is the inverse temperature and $V_{ij}^\text{max}$ is a constant equal to (or greater than) 
the maximum value of $V_{ij}$ over all $\bm{S}_i$ and $\bm{S}_j$. With this potential switching, 
the algorithm proceeds as follows:
\begin{itemize}
\item[(A)] Switch the potentials $V_{ij}$ to either $0$ or $\bar{V}_{ij}$ with the probability of $P_{ij}$ or $1-P_{ij}$, 
respectively.
\item[(B)] Perform a standard MC simulation with the switched Hamiltonian 
\begin{equation}
{\cal H}'=\sum\nolimits_{ij}'\bar{V}_{ij}(\bm{S}_i,\bm{S}_j),
\end{equation}
for $n_{\text{sw}}$ MC steps, where $\sum_{ij}'$ runs over all the potentials switched to $\bar{V}_{ij}$ and 
one MC step is defined by one trial for each $\bm{S}_i$ to be updated.
\item[(C)] Return to (A).
\end{itemize}
In the SCO method, an efficient method is employed to reduce the computational time of 
the potential switching in step (A) (see Ref.~\citen{SCO} for details). As a result, 
the computational time in step (A) becomes comparable to that in step (B) per MC step. 
For example, in the case of a two-dimensional magnetic dipolar system, both computational 
times are reduced to ${\cal O}(N)$.

\begin{figure}[t]
  \begin{center}
    \includegraphics[width=8cm]{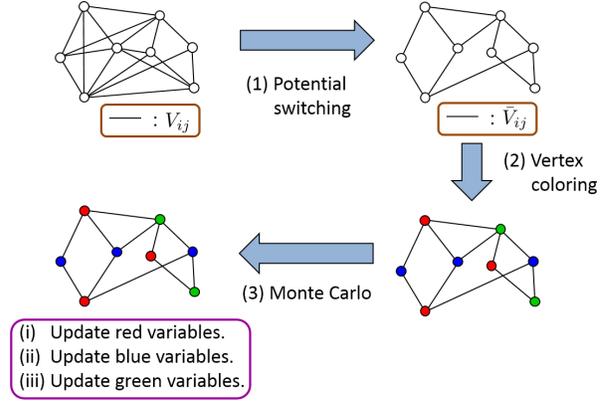}
    \caption{(Color online) Schematic illustration of parallel computations of the SCO method. 
      A vertex denotes a variable $\bm{S}_i$ and an edge denotes a potential 
      $V_{ij}$ or $\overline{V}_{ij}$. 
      In step (1), each potential is switched to either $0$ or $\overline{V}_{ij}$. 
      The edges whose potentials are switched to $0$ are eliminated in the subsequent steps. 
      In step (2), the vertex coloring of the graph is solved numerically in a parallel fashion. 
      In step (3), variables with a specific color are updated simultaneously by a
      standard MC simulation. This procedure is carried out for all the colors. 
    }
    \label{fig:outline_parallelSCO}
  \end{center}
\end{figure}

\subsection{Outline of parallel computations of the SCO method}

Figure~\ref{fig:outline_parallelSCO} shows a schematic illustration of parallel computations 
of the SCO method. A vertex denotes a variable $\bm{S}_i$ and an edge denotes 
a potential $V_{ij}$ or $\overline{V}_{ij}$. In step (1), each potential is switched to 
either $0$ or $\overline{V}_{ij}$. The edges whose potentials are switched to $0$ 
are eliminated in the subsequent steps. Because each potential is switched independently 
with the probability $P_{ij}$, we can easily parallelize the computation in this step. 
In step (2), the vertex coloring of the graph created in step (1) is solved 
numerically in a parallel fashion. The parallel computation of the vertex coloring 
will be explained in detail in the next subsection. Lastly, we perform a standard 
MC simulation with the switched Hamiltonian ${\cal H}'$ in step (3). It is apparent 
from the definition of the vertex coloring that variables with the same color 
do not interact with each other. Therefore, we can update variables with a specific 
color by parallel computation, as we do in MC simulations of Ising models 
by a checkerboard decomposition. By doing this simultaneous update for all the colors, 
we can parallelize the MC calculation in step (3). 

\subsection{Parallel computation of the vertex coloring}
\label{subsec:parallelColoring}

In this subsection, we briefly explain parallel computation of the vertex coloring. 
We refer the reader to the book in Ref.~\citen{book_KW} for more details. 
By solving the vertex coloring in a parallel fashion, we can perform all the three steps 
mentioned in the previous subsection by parallel computation. 
We hereafter call the vertex coloring by parallel computation {\it distributed graph coloring}.

The organization of this subsection is as follows: In Sect.~\ref{subsubsec:basis}, we explain 
the basis of the distributed graph coloring. In Sect.~\ref{subsubsec:BCR}, we explain 
a basic color reduction algorithm for the distributed graph coloring. 
This algorithm is used in an algorithm proposed by Kuhn and Wattenhofer,~\cite{KuhnWattenhofer06,book_KW} 
which is used in this study. This algorithm 
is explained in Sect.~\ref{subsubsec:KW}.

\subsubsection{Basis of the distributed graph coloring}
\label{subsubsec:basis}
We start with the introduction of several technical terms in the graph theory. The degree of a vertex is 
the number of edges that connect the vertex with others, and the maximum degree 
is the largest value of the degrees of a graph. In general, it is known that a graph 
with a maximum degree $\Delta$ can be colored with $\Delta+1$ colors, while, 
in most cases, it is not the smallest number of colors needed to color the graph. 
The aim of the distributed graph coloring is to color a graph with $\Delta+1$ colors 
by parallel computation.

In the distributed graph coloring, each vertex is initially colored by different colors, 
i.e., a graph is colored with $N$ colors, where $N$ is the number of vertices. 
The number of colors is gradually reduced from $N$ to $\Delta +1$ by repeating synchronous communication 
and parallel computation. In synchronous communication, vertices communicate with each other 
to know the colors of their neighbouring vertices. In parallel computation, all vertices 
simultaneously recolor themselves. The new color is locally calculated by using 
the information of the neighbouring colors obtained in the preceding communication. 
Vertices do not communicate with each other in parallel computation. 
The number of times of synchronous communications  required to accomplish a $(\Delta+1)$-coloring 
is called {\it running time}, hereafter denoted as $t_{\rm R}$. 
The main aim of the distributed graph coloring is to reduce $t_{\rm R}$ as much as possible.

\begin{figure}[t]
  \begin{center}
    \includegraphics[clip,width=8cm]{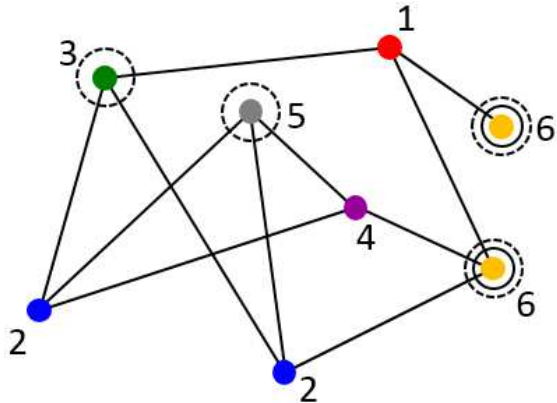}
    \caption{(Color online) A graph and its initial coloring to which we apply the BCR algorithm. 
      The number of vertices $N$ and the maximum degree $\Delta$ are $8$ and $3$, respectively. 
      The number of colors $\alpha$ is $6$. The vertices recolored in step (2) and those 
      recolored in step (2)' are enclosed by solid circles and dashed ones, respectively.} 
    \label{fig:BCR_UpdatedVertex}
  \end{center}
\end{figure}

\subsubsection{Basic color reduction algorithm}
\label{subsubsec:BCR}
The basic color reduction (BCR) algorithm is one of the most fundamental algorithms 
for the distributed graph coloring. Figure~\ref{fig:BCR_UpdatedVertex} shows a graph 
and its coloring to which we apply the BCR algorithm. We assume that the number of vertices 
and the maximum degree of the graph are $N$ and $\Delta$, respectively. 
The graph is initially colored with $\alpha$ colors $(\Delta+1 < \alpha \le N)$ and the coloring is valid, 
i.e.,  no adjacent vertices share the same color. The color of a vertex is specified 
by an integer between 1 and $\alpha$. 
In the BCR algorithm, the number of colors is reduced from $\alpha$ to $\alpha-1$ by the following steps:~\cite{book_DBR_BCR}

\begin{itemize}
\item [(1)] Each vertex communicates with each other to obtain the colors of the neighbouring vertices. 
\item [(2)] Each vertex recolors itself if its color is $\alpha$. The new color is chosen 
from a palette between $1$ and $\Delta+1$ by using the information obtained in step (1). 
\end{itemize}
Steps (1) and (2) correspond to synchronous communication and parallel computation
in the previous subsection, respectively. We can always choose a new color among $\Delta+1$ 
colors because the maximum degree of the graph is $\Delta$. It is also important to notice that 
the vertices with the color $\alpha$ cannot be adjacent to each other because the initial coloring 
is valid (see the vertices enclosed by a solid circle in Fig.~\ref{fig:BCR_UpdatedVertex}). 
This means that the new coloring is also valid even if each vertex with the color $\alpha$ 
simultaneously changes its color according to the information of the neighbouring colors. 
The BCR algorithm reduces the number of colors one at a time by repeating these two steps. 
Therefore, the running time $t_{\rm R}$ to accomplish a $(\Delta+1)$-coloring 
from an initial $N$-coloring is $N-\Delta-1$.

When we implemented the BCR algorithm in our simulation, we slightly modified the algorithm to improve its efficiency.
To be specific, we modified step (2) in the following manner:
\begin{itemize}
\item[(2)'] Each vertex recolors itself if its color is locally maximum. The new color is chosen 
from a palette between $1$ and $\Delta+1$ using the information obtained in step (1). 
\end{itemize}
In Fig.~\ref{fig:BCR_UpdatedVertex}, the vertices recolored in steps (2)' and (2) 
are enclosed by dashed circles and solid ones, respectively. We see that the former involves the latter. 
This means that the running time is reduced by this modification. We also find in Fig.~\ref{fig:BCR_UpdatedVertex} 
that the vertices recolored in step (2)' are not adjacent to each other because they are locally maximum. 
Therefore, the new coloring is also valid by the same reason as before. A demerit of this modification lies 
in the computational cost to check whether the color of a vertex is locally maximum. 
However, this demerit was not significant in our simulations because the degrees of graphs were not so large.

\begin{figure}[t]
  \begin{center}
    \includegraphics[clip,width=8cm]{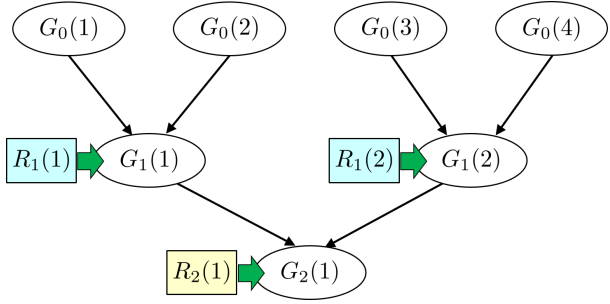}
    \caption{(Color online) Schematic illustration of the KW algorithm. 
      The number of vertices $N$ is related to the maximum degree $\Delta$ 
      by $N=(\Delta+1)\times 2^M$ with $M=2$.
      After two groups $G_1(1)$ and $G_1(2)$ are made by integrating two adjacent 
      groups at level $0$, we apply the color reductions $R_1(1)$ and $R_1(2)$ 
      to the groups $G_1(1)$ and $G_1(2)$, respectively. 
      These two color reductions are performed simultaneously. 
      The number of colors is halved from $4(\Delta+1)$ to 
      $2(\Delta+1)$ by the two color reductions. 
      We then apply the color reduction $R_2(1)$ to the group $G_2(1)$, 
      which is made by integrating the two groups $G_1(1)$ and $G_1(2)$. 
      As a result, the number of colors is reduced 
      from $2(\Delta+1)$ to $\Delta +1$.} 
    \label{fig:KWtree}
  \end{center}
\end{figure}

\subsubsection{KW algorithm}
\label{subsubsec:KW}
In this subsection, we explain an algorithm proposed by Kuhn and Wattenhofer.~\cite{KuhnWattenhofer06} 
We hereafter call it the KW algorithm. The KW algorithm markedly reduces the running time $t_{\rm R}$ 
by applying the BCR algorithm recursively. As mentioned above, we used this algorithm to numerically 
solve the vertex coloring.

Figure~\ref{fig:KWtree} shows a schematic illustration of the KW algorithm. 
For simplicity, we assume that the number of vertices $N$ and the maximum degree $\Delta$ 
are related by $N=(\Delta+1)\times 2^M$, where $M$ is an integer. 
Generalization to other cases is straightforward. We suppose that all the vertices are 
initially colored by different colors. The color is specified by an integer between $1$ and $N$. 
The KW algorithm starts by partitioning all vertices into $N/(\Delta+1)=2^M$ groups according to their colors. 
We hereafter denote them by $G_0(k)$ $(k=1,2,\cdots,2^{M})$, where the subscript ``0'' represents 
the level of the grouping. In this partitioning, vertices whose color is between $1+(k-1)(\Delta+1)$ and $k(\Delta+1)$ 
are assigned to the $k$-th group $G_{0}(k)$. We next make groups at level $1$ by integrating two adjacent groups 
at level $0$ (see Fig.~\ref{fig:KWtree}). We denote them by $G_1(k)$ $(k=1,2,\cdots,2^{M-1})$. 
Just after the integration, $2(\Delta+1)$ vertices in the group $G_1(k)$ are colored by $2(\Delta+1)$ colors. 
We then apply the BCR algorithm to reduce the number of colors from $2(\Delta+1)$ to $\Delta+1$. 
We denote this color reduction applied to vertices in the group $G_1(k)$ by $R_1(k)$. 
We can simultaneously perform all the color reductions at level $1$ because there is no overlap of colors 
among the groups. In the color reduction $R_1(k)$, the color of vertices is changed so that the new color 
is between $1+(k-1)(\Delta+1)$ and $k(\Delta+1)$. This guarantees that there is no overlap of colors 
among groups at level $1$ even after the color reductions. As a result of all the color reductions at level $1$, 
the number of colors used for coloring the whole graph is reduced from $N$ to $N/2$. 
By repeating the integration of two adjacent groups and the subsequent parallel color reduction $M$ times, 
the number of colors is reduced to $\Delta+1$.

We next consider the running time of the KW algorithm. 
At a level $p$, there are $2^{M-p}$ groups. 
In each group, the number of colors is reduced 
from $2(\Delta+1)$ to $\Delta+1$ by the BCR algorithm. Now, the point is that we can simultaneously 
perform BCRs in all the $2^{M-p}$ groups. To be specific, we simultaneously perform BCRs in all the groups 
to reduce the number of colors by one, and perform synchronous communication just once 
for the next color reductions. By repeating this procedure $\Delta+1$ times, 
we can reduce the numbers of colors of all the groups from $2(\Delta+1)$ to $\Delta+1$. 
The running time to achieve this color reduction at the level $p$ is $\Delta+1$. 
Because the number of levels is $M$, the total running time to reduce 
the number of colors from $N$ to $\Delta+1$ is estimated to be
\begin{equation}
t_{\rm R} = (\Delta +1) \times M = (\Delta +1)\log_2\left(\frac{N}{\Delta+1}\right),
\end{equation}
where we have used the relation $N=(\Delta+1)2^{M}$. If $N >> \Delta$, 
this running time is much shorter than 
that of the BCR algorithm, which is, as mentioned above, on the order of $N$.

\begin{figure}[t]
  \begin{center}
    \includegraphics[clip,width=8cm,angle=0]{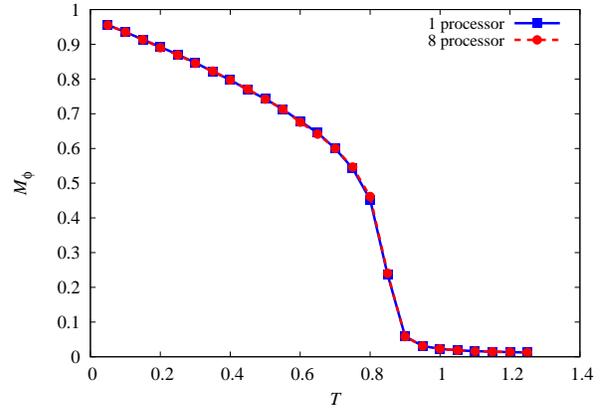}
    \caption{(Color online) The circular component of the magnetization $M_{\phi}$ 
      defined by Eq.~(\ref{eqn:Mphi}) is plotted as a function of $T/J$.
      The data for single-thread computation with $1$ processor 
      and those for multiple-thread computation with $8$ processors
      are denoted by squares and circles, respectively. 
      The size $L$ is $128$. The average is taken over $10$ 
      different runs with different initial conditions and 
      random sequences. }
    \label{fig:Mag_ave}
  \end{center}
\end{figure}

\section{Results}
\label{sec:result}
\subsection{Model}
To investigate the efficiency of parallel computation of the method developed in this study, 
we apply the method to a two-dimensional magnetic dipolar system on an $L\times L$ square lattice 
with open boundaries. The Hamiltonian of the system is described as
\begin{equation}
{\cal H}=-J\sum_{\langle ij\rangle}\bm{S}_i \cdot \bm{S}_j
+D\sum_{i<j}\left[ \frac{\bm{S}_i\cdot \bm{S}_j}{r_{ij}^3}
-3\frac{(\bm{S}_i\cdot\bm{r}_{ij})(\bm{S}_j\cdot\bm{r}_{ij})}
{r_{ij}^5}\right],
\label{eqn:Hamiltonian1}
\end{equation}
where $\bm{S}_i$ is a classical Heisenberg spin of $|\bm{S}_i|=1$, 
$\langle ij \rangle$ runs over all the nearest-neighbouring pairs, 
$\bm{r}_{ij}$ is the vector spanned from a site $i$ 
to $j$ in the unit of the lattice constant $a$, and $r_{ij}=|\bm{r}_{ij}|$. 
The first term describes short-range ferromagnetic exchange interactions and  
the second term describes long-range dipolar interactions, 
where $J(>0)$ is an exchange constant and $D(>0)$ is a constant 
that represents the strength of magnetic dipolar interactions. 
We hereafter consider the case that $D/J=0.1$. We choose this model 
because it was used as a benchmark of the SCO method.~\cite{SCO}
It is established that the model undergoes a phase transition from a paramagnetic state 
to a circularly ordered state at $T_{\rm c}\approx 0.88J$ as a consequence 
of the cooperation of exchange and dipolar interactions.~\cite{Sasaki96} 
We applied the SCO method only for magnetic dipolar interactions. 
The system was gradually cooled from an initial temperature $T=1.25J$ 
to $0.05J$ in steps of $\Delta T=0.05J$. The initial temperature was 
set to be well above the critical temperature. 
We set $n_{\rm sw}$ defined in Sect.~\ref{subsec:SCO} to be $100$, 
i.e., potential switching and subsequent vertex coloring are 
performed every $100$ MC steps. It was determined in Ref.~\citen{SCO} that 
this frequency of potential switching is sufficient for this model 
to obtain reliable results.

To check the correctness of our parallel computation, we performed MC simulation 
and measured the absolute value of the circular component of magnetization defined by
\begin{equation}
M_{\phi} \equiv \left\langle \left| \left[ \frac{1}{N} \sum_{i=1}^{N} 
\bm{S}_i \times \frac{\bm{r}_i-\bm{r}_{\rm c}}
{|\bm{r}_i-\bm{r}_{\rm c}|}
\right]_z \right|\right\rangle,
\label{eqn:Mphi}
\end{equation}
where $[ \cdots ]_z$ denotes the $z$-component of a vector, 
$\langle \cdots\rangle$ denotes thermal average, 
and $\bm{r}_{\rm c}$ is a vector describing the center 
of the lattice. In this measurement, the system was kept 
at each temperature for $100,000$ MC steps. 
The first $50,000$ MC steps are for equilibration and the following 
$50,000$ MC steps are for measurement. 
We performed simulations for $10$ different runs 
with different initial conditions and random sequences. 
The result is shown in~Fig.~\ref{fig:Mag_ave}. The squares and circles denote 
the result of single-thread computation with $1$ processor and 
that of multiple-thread computation with $8$ processors, respectively. 
Both data coincide with each other within statistical error. 
We also see that $M_\phi$ rapidly increases 
around the critical temperature $T_{\rm c}\approx 0.88J$. From these results, 
we conclude that our parallel computation is performed correctly.


\begin{table}[t]
\caption{Temperature dependences of the mean degree $\langle k \rangle$ and the maximum degree 
$\langle \Delta \rangle$. Graphs are created by the potential switching in the SCO method. 
The size $L$ is 2304. The average is taken over $35$ graphs. 
}
\label{table:degree_mean_max}
\begin{tabular}{ccc}
\hline
temperature & $\langle k \rangle$ & $\langle \Delta \rangle$ \\
\hline
$1.25J$ & 1.40 & 8.54\\
$0.45J$ & 3.36 & 12.9\\
$0.05J$ & 22.7 & 41.6\\
\hline
\end{tabular}
\end{table}

\begin{figure}[t]
  \begin{center}
    \includegraphics[clip,width=8cm,angle=0]{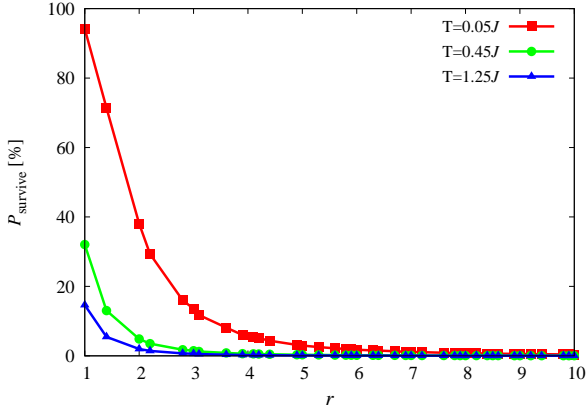}
    \caption{(Color online) The probability $P_{\rm survive}$ that a potential survives 
      by being switched to ${\bar V}$ is plotted as a function of $r$, 
      where $r$ is the distance between two interacting sites. 
      \textcolor{black}{The size $L$ is $2304$.} The temperatures are $0.05J$ (squares), 
      $0.45J$ (circles), and $1.25J$ (triangles), respectively.
    }
    \label{fig:Psurvive_vs_R}
  \end{center}
\end{figure}

\begin{figure}[t]
  \begin{center}
    \includegraphics[clip,width=8cm]{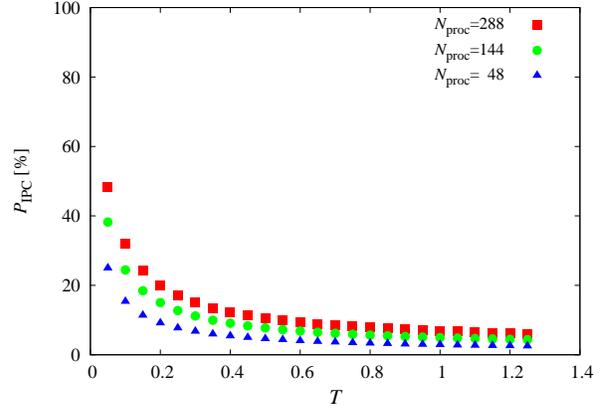}
    \caption{(Color online) The proportion $P_{\rm IPC}$ of surviving potentials that require 
      interprocessor communication is plotted as a function of temperature. 
      The size $L$ is $2304$. The numbers of processors $N_{\rm proc}$'s are $48$ (triangles), 
      $144$ (circles), and $288$ (squares), respectively. 
      
    }
    \label{fig:Ratio_IPC}
  \end{center}
\end{figure}

\subsection{Properties of graphs and improvements to reduce communication traffic}
\label{subsec:improvement}
In Table~\ref{table:degree_mean_max}, we show the mean degree $\langle k \rangle$ and the maximum degree 
$\langle \Delta \rangle$ of graphs at three temperatures. These are important quantities because 
the maximum degree determines the number of colors and the mean degree $\langle k \rangle$ is 
proportional to the computational time per MC step. The size $L$ is $2304$. 
As found in Ref.~\citen{SCO}, these quantities hardly depend on the size 
in two-dimensional magnetic dipolar systems if the size is sufficiently large. 
As expected from Eq.~(\ref{eqn:Pij}), both $\langle k \rangle$ and $\langle \Delta \rangle$ increase 
with decreasing temperature. However, they are several tens at most. This means that 
most of the interactions are cut off by the potential switching. It should be noted 
that both $\langle k \rangle$ and $\langle \Delta \rangle$ are $N-1\approx 5\times 10^6$ 
before potentials are switched. Figure~\ref{fig:Psurvive_vs_R} shows the distance 
dependence of the probability $P_{\rm survive}$ that a potential survives 
by being switched to $\bar V$. The temperatures are the same as those 
in Table~\ref{table:degree_mean_max}. We see that the probability increases with 
decreasing temperature. The probability is close to one when $T=0.05J$ and $r=1$. 
However, $P_{\rm survive}$ rapidly decreases with increasing $r$ at any temperature.

Taking these properties of the SCO method into consideration, we implemented our simulation 
in the following way: We first divide a lattice into $N_{\rm proc}$ square cells 
($N_{\rm proc}$ is the number of processors) and assign each cell to each processor. 
We then list the vertices whose information should be sent by interprocessor communication 
when we update spins with a certain color. For example, a vertex $i$ is added to a list 
for red-spin update if it satisfies the following two conditions: 
\begin{itemize}
\item[$\bullet$] The vertex $i$ is connected with a red vertex $j$. 
\item[$\bullet$] The two vertices $i$ and $j$ belong to different cells. 
\end{itemize}
This list is made for each color just once when a new graph is created by the potential switching. 
When we update spins of a certain color, we perform interprocessor communication in advance 
according to the list. Although it requires some computational cost to make the lists, 
they enable us to reduce the communication traffic before 
parallel MC calculation as much as possible. Figure~\ref{fig:Ratio_IPC} shows 
the temperature dependence of the proportion \textcolor{black}{$P_{\rm IPC}$} 
of surviving potentials that require 
interprocessor communication. The proportion increases with increasing number 
of processors. Note that the mesh size decreases as the number 
of processors increases. The proportion also increases with decreasing temperature. 
However, it is less than $20\%$ in most cases, meaning that communication traffic 
is considerably reduced by this improvement.

\begin{figure}[t]
  \begin{center}
    \includegraphics[clip,width=8cm,angle=0]{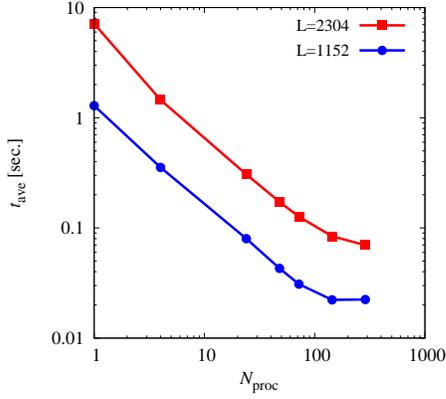}
    \caption{\textcolor{black}{(Color online) The average computational time $t_{\rm ave}$ in the strong scaling 
      is plotted as a function of the number of processors $N_{\rm proc}$ 
      for $L=2304$ (squares) and $L=1152$ (circles). 
      $t_{\rm ave}$ is defined by Eq.~(\ref{eqn:t_ave}).} 
      The average is taken over the temperatures between $0.05J$ and $1.25J$.
    }
    \label{fig:Time_vs_CPU}
  \end{center}
\end{figure}

\begin{figure}[t]
  \begin{center}
    \includegraphics[width=8cm,angle=0]{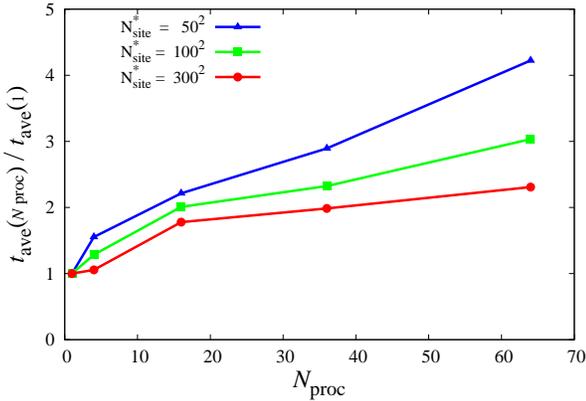}
    \caption{\textcolor{black}{(Color online) The ratio $t_{\rm ave}(N_{\rm proc})/t_{\rm ave}(1)$ in the weak scaling is plotted 
    as a function of $N_{\rm proc}$, where $t_{\rm ave}(X)$ is the average computational time 
    defined by Eq.~(\ref{eqn:t_ave}) when the number of processors is $X$. 
      The numbers of sites per processor $N_{\rm site}^*$'s are 
      $50^2$ (triangles), $100^2$ (squares), and $300^2$ (circles), respectively. 
      The average is taken over the temperatures between $0.05J$ and $1.25J$.}}
    \label{fig:weakscaling_rate}
  \end{center}
\end{figure}

\subsection{Efficiency of parallel computation}
In Fig.~\ref{fig:Time_vs_CPU}, we plot the average computational time per MC step $t_{\rm ave}$ 
as a function of the number of processors. The average time $t_{\rm ave}$ is defined by
\begin{equation}
t_{\rm ave}=\frac{1}{100}t_{\rm switch}+\frac{1}{100}t_{\rm color}+t_{\rm MC},
\label{eqn:t_ave}
\end{equation}
where $t_{\rm switch}$, $t_{\rm color}$, and $t_{\rm MC}$ are the computational times 
to switch potentials, to solve vertex coloring, and to perform MC simulation for one MC step, 
respectively. Recall that potential switching and the subsequent vertex coloring are performed 
every $100$ MC steps. The average is taken over the temperatures between $0.05J$ and $1.25J$. 
The data for $L=2304$ and those for $L=1152$ are denoted by squares and circles, respectively. 
\textcolor{black}{The data correspond 
to the strong scaling because $N_{\rm proc}$ is increased with the system size $L$ fixed.} 
When $L$ is $2304$, the computations with $144$ and $288$ processors 
are about $85$ and $102$ times faster than that with one processor, respectively. 
In the case of $L=1152$, the speedups by $144$ and $288$ processors 
are about $58$ and $57$, respectively. 

\textcolor{black}{
Figure~\ref{fig:weakscaling_rate} shows the data of the weak scaling.
In the weak scaling, we increase both the system size and $N_{\rm proc}$ 
with the number of sites per processor $N_{\rm site}^*$ fixed. 
In the figure, the ratio $t_{\rm ave}(N_{\rm proc})/t_{\rm ave}(1)$ is 
plotted as a function of $N_{\rm proc}$, where $t_{\rm ave}(X)$ is the average 
computational time when the number of processors is $X$. 
We see that the ratio decreases with increasing $N_{\rm site}^*$. 
This means that the parallelization efficiency is improved as the system size increases. 
}

\begin{figure}[t]
  \begin{center}
    \includegraphics[width=8cm,angle=0]{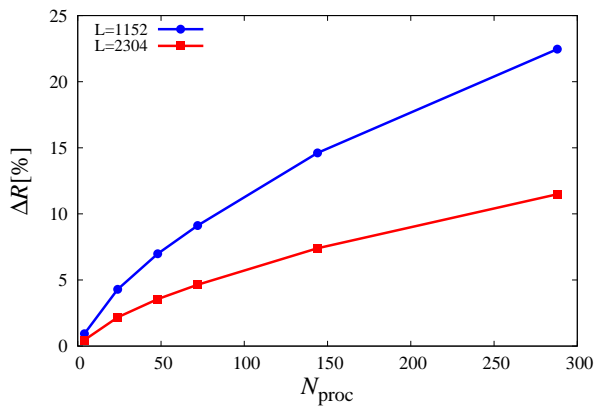}
    \caption{\textcolor{black}{(Color online) $\Delta R$ defined by Eq.~(\ref{eqn:def_DeltaR}) is plotted as a function of 
    $N_{\rm proc}$ for $L=1152$ (circles) and $L=2304$ (squares). The average is taken over 
    the colorings and the colors of each coloring. The number of colorings is 10.
    The temperature $T$ is $0.05J$. 
     }}
    \label{fig:how_much_color}
  \end{center}
\end{figure}

\textcolor{black}{
We next consider fluctuations in the number of sites assigned to each processor 
in the MC calculation. As mentioned in Sect.~\ref{subsec:improvement}, 
when we update spins with some color in the MC calculation, 
each processor updates spins in the assigned cell. The number of sites 
with the color is different from cell to cell. Because these fluctuations cause 
the difference in the computational time among processors, it may significantly 
decrease the parallelization efficiency. To evaluate the effect of fluctuations, 
we measured the following quantity:
\begin{equation}
\Delta R \equiv \frac{K_{\rm max}-K_{\rm ave}}{K_{\rm ave}},
\label{eqn:def_DeltaR}
\end{equation}
where $K_{\rm max}$ and $K_{\rm ave}$ are the maximum and average values 
of the number of sites with a certain color, respectively. We measured this quantity because 
the computational time is determined by the maximum number of sites 
among processors. Figure~\ref{fig:how_much_color} shows 
the $N_{\rm proc}$ dependence of $\Delta R$. The average is taken over 
the colorings and the colors of each coloring. The temperature $T$ is $0.05J$. 
Because the number of colors increases with decreasing temperature, 
this temperature corresponds to the worst case. 
The data for $L=1152$ and those for $L=2304$ are denoted by circles and squares, respectively. 
When $N=1152$ and $N_{\rm proc}=288$, $\Delta R$ is about $22\%$. This means that 
the fluctuations decrease the parallelization efficiency to some extent. 
However, we also see that the fluctuations decrease with increasing system size. 
}

\begin{figure}[t]
  \begin{center}
    \includegraphics[clip,width=8cm,angle=0]{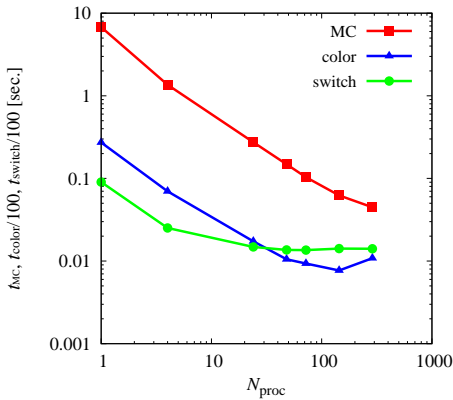}
    \caption{(Color online) $t_{\rm MC}$ (squares), $t_{\rm color}/100$ (triangles), and $t_{\rm switch}/100$ (circles) 
      are plotted as functions of the number of processors $N_{\rm proc}$. 
      The size $L$ is $2304$. The average is taken over the temperatures between 
      $0.05J$ and $1.25J$. 
    }
    \label{fig:Breakdown}
  \end{center}
\end{figure}

In Fig.~\ref{fig:Breakdown}, $t_{\rm switch}/100$, $t_{\rm color}/100$, 
and $t_{\rm MC}$ are plotted as functions of the number of processors. 
The size $L$ is $2304$. The sum of the three computational times is equal to 
$t_{\rm ave}$ for $L=2304$ shown in Fig.~\ref{fig:Time_vs_CPU} (see Eq.~(\ref{eqn:t_ave})). 
The computation time of MC simulation $t_{\rm MC}$ is dominant owing to the factor $1/100$
in $t_{\rm switch}$ and $t_{\rm color}$. \textcolor{black}{ 
We see that each computational time saturates as $N_{\rm proc}$ increases. 
Although there are several causes of the saturation, such as the fluctuations 
in the number of sites discussed in the previous paragraph, the main reason 
for the saturation is an increase in communication traffic. In the
MC calculation, we reduced communication traffic by the method 
described in Sect.~\ref{subsec:improvement}.
Therefore, the saturation of $t_{\rm MC}$ is gradual. In contrast, the proportion 
of communication traffic is large in the potential switching and coloring 
because the improvement mentioned in Sect.~\ref{subsec:improvement} 
is not applicable to them.}
To make the present method effective even for larger parallel computations, 
we need to improve the parallelization efficiencies of the two processes.

\section{Conclusions}
\label{sec:conclusions}
In this study, we have developed a method of parallelizing the SCO method, 
which is a MC method for long-range interacting systems. 
To parallelize the MC calculation in the SCO method, we numerically solve the vertex 
coloring of a graph created by the SCO method. This computation is performed 
in parallel using the KW algorithm.~\cite{KuhnWattenhofer06,book_KW}
We applied this method to a two-dimensional magnetic dipolar system 
on an $L\times L$ square lattice to examine its parallelization efficiency. 
The result showed that, in the case of $L=2304$, the speed of computation 
increased about $102$ times by parallel computation with $288$ processors. 

\section*{Acknowledgments}
This work was supported by JSPS KAKENHI Grant Number 25400387. Some of 
the experimental results in this research were obtained using supercomputing 
resources at Cyberscience Center, Tohoku University.


\end{document}